
\magnification=\magstep1
\baselineskip=15pt
\overfullrule=0pt
\nopagenumbers
\font\twelvebf=cmbx12

\def\z{\bar{z}}
\def\w{\bar{w}}

\rightline{hep-th/9211070}
\rightline{IASSNS-HEP-92/34}
\rightline{Oct 1992}
\vskip .7in
\centerline{\twelvebf Self-Dual Yang-Mills}
\centerline{\twelvebf and the}
\centerline{\twelvebf Hamiltonian Structures of Integrable Systems}
\vskip 1in
\centerline{ Jeremy Schiff}
\smallskip
\centerline{\it Institute For Advanced Study}
\centerline{\it Olden Lane, Princeton, NJ 08540}
\vskip 1in
\centerline{\bf Abstract}
\smallskip
\noindent
In recent years it has been shown that many, and possibly all, integrable
systems can be obtained by dimensional reduction of self-dual Yang-Mills.
I show how the integrable systems obtained this way naturally inherit
bihamiltonian structure. I also present a simple,
gauge-invariant formulation of the self-dual Yang-Mills hierarchy proposed
by several authors, and I discuss the notion of gauge equivalence of
integrable systems that arises from the gauge invariance of the self-duality
equations (and their hierarchy); this notion of gauge equivalence may well
be large enough to unify the many diverse existing notions.
\vfill\eject
\footline={\hss\tenrm\folio\hss}
\pageno=1

\noindent{\it Contents}
\smallskip
\noindent
\item{1.} Introduction
\item{2.} The Self-Dual Yang-Mills Equations and Hierarchy
\item{3.} Gauge Choices
\item{4.} Bihamiltonian Integrable Systems
\item{5.} Dimensional Reduction of SDYM
\item{6.} Example: KdV
\item{7.} Some Formal Manipulations with Formulae of Drinfeld and Sokolov
\item{8.} The Bihamiltonian Structure Induced on Reductions of SDYM
\item{9.} Example: The Heisenberg Ferromagnet
\item{10.} The KP Hierarchy from the SDYM Hierarchy
\item{11.} Conclusions and Further Directions
\item{12.}   Acknowledgements
\item{13.}   References

\vskip.2in

\noindent{\it 1. Introduction}
\smallskip
\noindent Early on in the study of the self-dual Yang-Mills (SDYM)
equations it was observed that dimensional reductions of these
equations give rise to so-called ``integrable systems'' [1].
It was conjectured by Ward [2] that SDYM may be a {\it universal}
integrable system, i.e. that {\it all} integrable systems might be
obtained from it by suitable reductions. A remarkable piece of evidence
for this was produced a few years ago by Mason and Sparling [3], who showed
how to obtain the Korteweg-de Vries (KdV) and nonlinear Schr\"odinger (NLS)
equations from SDYM, and further wrote down a hierarchy of gauge-invariant
equations from which the KdV and NLS hierarchies could be obtained.
This stimulated much activity. Mason and Sparling just examined SDYM with
gauge group $SL(2)$; Bakas and Depireux showed how to obtain certain flows
in the $SL(N)$-KdV hierarchy from $SL(N)$ SDYM, and also obtained a
apparently new hierarchy from $SL(3)$ SDYM, with a hamiltonian structure
given by the classical limit of the $W_3^2$ conformal algebra [4]. I showed
how the reduction method of Mason and Sparling could be extended to
obtain certain three dimensional versions of the KdV and NLS equations
{}from SDYM [5]; this was also observed by Strachan, who studied the
resulting equations in some depth [6]. In a recent paper Ablowitz
{\it et.al.} introduced an apparently different version of the SDYM
hierarchy, and showed, amongst other things, how the KP hierarchy can
be obtained from SDYM with an infinite-dimensional gauge group [7].

Certain well-known properties of integrable systems are very
obviously inherited from SDYM. Amongst these are the Lax pair formulation,
given for SDYM by Belavin and Zakharov [8], and the Painlev\'e property [9],
shown for SDYM in [10]. Other properties of integrable systems have
at least natural analogs in SDYM. We might expect the inverse scattering
formalism to be a special case of the twistor formalism
for SDYM, and this is indeed the case [3]; we might expect B\"acklund
transformations for integrable systems to originate in those for SDYM [11],
or maybe in the loop group action on solutions of SDYM [12];
finally, for certain integrable systems there exist direct solution
methods (``the Hirota method''), and these presumably  arise in cases
where we can actually solve the inverse twistor transform and thereby
write down solutions exploiting the action of the loop group on the twistor
data [12] (the ``Hirota variables'', which have the property
of being entire [13], are presumably related to the patching matrix $G$ in
[12]; I draw the reader's attention to a recent work on the relation of
the zero-curvature and Hirota formulations of integrable systems [14], which
might be exploited to decide these issues).

None of the above issues will be treated further in this paper.
Here I wish to devote myself to one property of integrable systems which
has an obvious analog in the SDYM equations, and another that does not.
There is a notion of ``gauge equivalence'' between different integrable
systems, two systems being said to be equivalent when there is a map
{}from the solutions of one to the solutions of the other. The classic
example of such a map is the Miura map between solutions of MKdV and KdV.
These maps arise in many diverse settings. The origin of these maps at
the level of SDYM is just gauge freedom in the SDYM
equations; performing the same reduction in different gauges gives
rise to different equations. I will give a detailed analysis of
different gauges in which one might consider reduction of SDYM, and show
how the usual notions of gauge equivalence (for instance that of
Drinfeld and Sokolov [15]) emerge.

The property of integrable systems I wish to explain that does not
have an immediate analog in SDYM is that of bihamiltonian structure.
It turns out that this has its origins in the fact
that  the space of solutions of SDYM has three
symplectic structures. These are usually thought of as gauge invariant,
but this assumes certain boundary conditions; for the types
of solution we will be looking at, we will see the symplectic structures
have certain restricted gauge invariances, and they give (on reduction)
the bihamiltonian structures of integrable systems. This is the main
result of this paper; understanding different gauge choices etc. is a
necessary tool for this analysis.

The contents of this paper have been given above. While not strictly
necessary for the results on hamiltonian structures, I have included in
section 2 what I consider to be the most natural definition of the SDYM
hierarchy, and I show in section 3 that it contains the hierarchies of
[3], [6] and [7]; in section [11] I give a reduction of the hierarchy to the
KP hierarchy (c.f.[7]).

\vskip.2in

\noindent{\it 2. The Self-Dual Yang-Mills Equations and Hierarchy}
\smallskip
\noindent The SDYM equations on ${\bf R}^4$ with coordinates $z,\z,w,\w$
and metric $ds^2=dwd\z-dzd\w$ are (in standard notation)
$$ F_{zw}=0  \eqno{(1a)}$$
$$ F_{\z\w}=0  \eqno{(1b)}$$
$$ F_{z\w}-F_{w\z} =0  \eqno{(1c)}$$
These are consistency conditions for the linear system
$$ \eqalign{ (D_z+\lambda D_{\z})\Psi &=0 \cr
             (D_w+\lambda D_{\w})\Psi &=0 \cr} \eqno{(2)}$$
There is a natural generalization of the above equations we might write down.
Let $z_1,...,z_n,\z_1,...,\z_n$ be coordinates for ${\bf R}^{2n}$, $n\ge 2$;
the consistency conditions for the linear system
$$ (D_{z_i}+\lambda D_{\z_i})\Psi =0,~~~~~~~~~~~~i=1,...,n   \eqno{(3)}$$
are
$$ \eqalign{  F_{z_iz_j}&=0  \cr
            F_{\z_i\z_j}&=0~~~~~~~~~~~~~~~i,j\in\{1,...,n\};i\not=j  \cr
            F_{z_i\z_j}-F_{z_j\z_i}&=0  \cr }\eqno{(4)}$$
If $d$ is the dimension of the gauge group, then there are ${3\over 2}dn(n-1)$
equations here for $d(2n-1)$ unknowns (components of the gauge field modulo
gauge transformations); the equations are therefore overdetermined for $n>2$.
For $n=4$ equations (4) are equations (3.5)$'$ in [16]. I will take equations
(4) with $n=\infty$ as my definition of the SDYM hierarchy. Whereas equations
(1) can be modeled on any manifold with a metric, I cannot see how to
write equations (4) on anything more general than the cartesian product
of $2n$ 1-dimensional manifolds.

\vskip.2in

\noindent{\it 3. Gauge Choices}
\smallskip
\noindent There are two ways to approach solving equations (1) I wish to
consider, and these extend naturally to the SDYM hierarchy. Yang [17]
proposed first solving equations (1a) and (1b) to write
$$ \eqalign{ A_z&=g^{-1}\partial_z g \cr
             A_w&=g^{-1}\partial_w g \cr }\eqno{(5a)}$$
$$ \eqalign{ A_{\z}&=h^{-1}\partial_{\z} h \cr
             A_{\w}&=h^{-1}\partial_{\w} h \cr }\eqno{(5b)}$$
where $g,h$ are some gauge group valued functions. Gauge transformations
($A\rightarrow\Lambda^{-1}A\Lambda + \Lambda^{-1}d\Lambda$)   act  on $g,h$
by $g\rightarrow g\Lambda$, $h\rightarrow h\Lambda$. Equation (1c) reduces to
$$ \partial_z(J^{-1}\partial_{\w}J)-\partial_w(J^{-1}\partial_{\z}J)=0
                      \eqno{(6)}$$
where $J=hg^{-1}$ (which is gauge invariant); equation (6) can also be written
$$ \partial_{\z}(J^{-1}\partial_wJ)-\partial_{\w}(J^{-1}\partial_zJ)=0
                      \eqno{(6)'}$$
Equation (6) or (6)$'$ is known as the $J$ formulation of SDYM.
The other approach to SDYM (which I believe was first given implicitly in
[18]) is to solve (1b) to express $A_{\z},A_{\w}$ in terms of $h$ (equation 5b)
and then to solve (1c) to get
$$ \eqalign{ A_z&=h^{-1}\partial_z h + h^{-1}\partial_{\z}N h \cr
             A_w&=h^{-1}\partial_w h + h^{-1}\partial_{\w}N h \cr} \eqno{(7)}$$
where $N$ is some gauge-invariant function valued in the Lie algebra of the
gauge group. (1a) then yields
$$ (\partial_w\partial_{\z}-\partial_z\partial_{\w})N +
   [\partial_{\w}N,\partial_{\z}N] =0  \eqno{(8)}$$
If we write $M=\partial_{\z}N$ then we can write this, at least formally, as
$$ \partial_wM=(\partial_z + [M,~~])\partial_{\z}^{-1}\partial_{\w}M
        \eqno{(9)}$$
I call this the $M$ formulation of SDYM. Treating $w$ as ``time'', this
has the form of an evolution equation. There is a clear hint from (9) that
if we wish to obtain local evolution equations by reduction of SDYM, what
we need to do is fix the $\z$ dependence so that the $\partial_{\z}^{-1}$
integration symbol can be integrated out. Note that a lagrangian for (8)
was given in [19], and one for (6) was given in [20].

In [18] it was noted that there exists an integrable hierarchy of which
(9) is the first non-trivial member,
$$ \partial_{z_i}M = [(\partial_z + [M,~~])\partial_{\z}^{-1}]^{i-1}
     \partial_{\w}M~~~~~~~~~~i=2,3,... \eqno{(10)}$$
Let us consider our SDYM hierarchy (4). Solving the second and third sets
of equations we have
$$ \eqalign{A_{\z_i}&=h^{-1}\partial_{\z_i} h \cr
            A_{z_i} &=h^{-1}\partial_{z_i}  h + h^{-1}\partial_{\z_i}N h \cr}
                        \eqno{(11)}$$
Write $M=\partial_{\z_1}N$. The $j=1$ equations in the first set of equations
of (4) give
$$ \partial_{z_i}M = (\partial_{z_1} + [M,~~])\partial_{\z_1}^{-1}
     \partial_{\z_i}M~~~~~~~~~~i=2,3,... \eqno{(12)}$$
Writing $z_1=z,\z_1=\z,\z_2=\w$ it is clear that if we impose the dimensional
reductions $\partial_{\z_i}=\partial_{z_{i-1}}$, $i=3,4,...$ on (12), then we
recover (10). This latter hierarchy is essentially a dimensionally reduced
version of our hierarchy in $A_{\z_i}=0$ gauge. The hierarchy of [7]
(equation (21)) is
a slightly generalized version of this, where the $A_{\z_i}$'s are allowed
to be arbitrary commuting constant matrices (though the authors are clearly
quite aware that one can generate more general possibilities via gauge
transformation). The Bogomolnyi hierarchy of [3], which is written in
a gauge invariant way, can be obtained from ours by the dimensional
reduction $\partial_{\z_i}=\partial_{z_{i-1}}$, $i=1,2,...$ (so
$\partial_{\z_1}=0$). The hierarchy of [6] is a generalization of this,
obtained from ours by the dimensional
reduction $\partial_{\z_i}=\partial_{z_{i-1}}$, $i=1,2,...$, $i\not= m$,
where $m$ is some integer greater than $1$.
I have not investigated the relationship of the
hierarchy (4) with that of [21]. Especially for the purpose of discussing
reductions of SDYM, where, as I mentioned in the introduction, reductions
in different gauges give rise to equations related by Miura maps, it is
important to have the full gauge-invariant version of the hierarchy.

Finally in this section, we will need later the version of $M$ formulation
where we first solve (1a) and (1c). Solving (1a) gives us (5a), and solving
(1c) gives
$$ \eqalign{ A_{\z}&=g^{-1}\partial_{\z}g + g^{-1}\partial_zPg \cr
             A_{\w}&=g^{-1}\partial_{\w}g + g^{-1}\partial_wPg \cr}
           \eqno{(13)}$$
Equation (1b) now gives
$$ (\partial_{\w}\partial_z-\partial_{\z}\partial_w)P + [\partial_wP,
        \partial_zP]=0   \eqno{(14)}$$

\vskip.2in

\noindent{\it 4. Bihamiltonian Integrable Systems} [22]
\smallskip
\noindent
An evolution equation has (local) bihamiltonian form
when it can be written in the form
$$ {{\partial{\bf u}}\over{\partial w}}
   ={\cal D}_1{{\delta{\cal H}_3[{\bf u}]}\over{\delta{\bf u}}}
   ={\cal D}_2{{\delta{\cal H}_2[{\bf u}]}\over{\delta{\bf u}}}
         \eqno{(15)}$$
where ${\cal D}_1,{\cal D}_2$ are (local) coordinated hamiltonian operators
and ${\cal H}_2[{\bf u}],{\cal H}_3[{\bf u}]$ are suitable
functionals of ${\bf u}(x)$.
The recursion operator is defined by ${\cal R}={\cal D}_2{\cal D}_1^{-1}$.
The conserved quantities are given recursively by
$$ {\cal D}_1{{\delta{\cal H}_i}\over{\delta{\bf u}}}
 = {\cal D}_2{{\delta{\cal H}_{i-1}}\over{\delta{\bf u}}}  \eqno{(16)}$$
and the associated hierarchy of equations is given by
$$ {{\partial{\bf u}}\over{\partial z_i}}
                ={\cal D}_1{{\delta{\cal H}_{i+1}}\over{\delta{\bf u}}}
                ={\cal D}_2{{\delta{\cal H}_i}\over{\delta{\bf u}}}
           ={\cal R}^{i-2}{\cal D}_2{{\delta{\cal H}_2}\over{\delta{\bf u}}},
         ~~~~~~~~~i=2,3,...                  \eqno{(17)}$$
The classic example is the KdV equation for which
$$ \eqalign{ {\cal D}_1&=\partial_x \cr
             {\cal D}_2&=-\partial_x^3+u(x)\partial_x+\partial_xu(x) }$$
$$ \eqalign{ {\cal H}_2&=\int dx~{\textstyle{1\over 2}}u^2 \cr
             {\cal H}_3&=\int dx~{\textstyle{1\over 2}}(u_x^2 +u^3)
                                  }\eqno{(18)}$$
Without going into details we note the existence of ``elementary dimensional
deformations'' [23] of bihamiltonian integrable systems. For example,
the KdV equation can
be written $u_w={\cal R}u_x$, and the equation $u_w={\cal R}u_y$ describing
the evolution in $w$ of a function $u(x,y,w)$ is also integrable. This
equation is non-local, but by the substitution $u=\gamma_x$ is made local;
it has bihamiltonian form (1) with the same ${\cal D}_1,{\cal D}_2$ as
for the KdV equation but with
$$ \eqalign{ {\cal H}_2&=\int dxdy~{\textstyle{1\over 2}}\gamma_x\gamma_y \cr
      {\cal H}_3&=\int dxdy~{\textstyle{1\over 2}}(\gamma_{xx}\gamma_{xy} +
                            \gamma_x^2\gamma_y) }
              \eqno{(19)}$$

SDYM in $M$ formulation is a bihamiltonian integrable system; on the space
of Lie algebra valued functions $M(z,\z)$, ${\cal D}_2=\partial_z+[M,~~]$
and ${\cal D}_1=\partial_{\z}$ are two coordinated hamiltonian operators.
We clearly have
$$ {\cal H}_2={\textstyle{1\over 2}}\int dzd\z d\w ~Tr(M\partial_{\z}^{-1}
                       \partial_{\w}M) \eqno{(20)}$$
but it is not so easy to write expressions for the higher conserved quantities.
Note though we have $\partial_{z_i}M=(\partial_z+[M,~~])\delta{\cal H}_i/
\delta M$. This equation is central in Chern-Simons theory; using
$M=-\partial_zJJ^{-1}$ we deduce we can write
$$ {\cal H}_n=\int d\z d\w~S^{(i)}_{WZW}[J]  \eqno{(21)}$$
where $S^{(i)}_{WZW}$ denotes the WZW action on the plane defined by
coordinates $z,z_i$. This satisfies (up to an irrelevant overall factor)
$$ \delta S^{(i)}_{WZW}=\int dzdz_n~Tr[J^{-1}\delta J \partial_z(J^{-1}
      \partial_{z_i}J)]   \eqno{(22)}$$
(see [24]), from which it is easy to check that ${\cal H}_i$ as given in
(21) is independent of all the times $z_j$, $j\ge 2$ (by construction it
is independent of $z_i$). It is not clear to me at the moment whether
there are analogs of ${\cal H}_i$ for the more general hierarchy (4).

\vskip.2in

\noindent{\it 5. Dimensional Reduction of SDYM}
\smallskip
\noindent Following [3], let us consider reducing (1) by requiring the
potentials to be independent of $\z$. This is not  a gauge invariant statement;
to avoid problems we restrict ourselves to $\z$-independent gauge
transformations. Since $A_{\z}$ transforms homogeneously under such gauge
transformations, $A_{\z}\rightarrow\Lambda^{-1}A_{\z}\Lambda$, we can no
longer fix the gauge $A_{\z}=0$. Restricting ourselves further to the
case where $A_{\z}$ in some gauge is constant, for $SL(2)$ the equivalence
classes of such $A_{\z}$'s under gauge transformations are represented by
$$ \pmatrix{0&0\cr1&0\cr},~~~ \kappa\pmatrix{1&0\cr0&-1\cr} \eqno{(23)}$$
where $\kappa$ is an arbitrary constant.
These give rise to integrable systems related to KdV and NLS respectively
[3]. For $SL(N)$, the class represented by $A_{\z}$ with $(A_{\z})_{ij}=
\delta_{iN}\delta_{j1}$ gives rise to the SL(N)-KdV equation, and the class
represented by $A_{\z}={\rm diag}(\kappa_1,...,\kappa_N)$,
$\kappa_1+...+\kappa_N=0$, gives rise to the generalized NLS equation of
Fordy and Kulish [25]. For $SL(3)$ the class
$$ A_{\z}=\pmatrix{0&0&0\cr
                   1&0&0\cr
                   0&1&0\cr}  \eqno{(24)}$$
gives rise to the $W_3^2$ hierarchy of Bakas and Depireux [4], and the
remaining class
$$ A_{\z}=\pmatrix{\kappa&0&0\cr
                   1&\kappa&0\cr
                   0&0&-2\kappa\cr} \eqno{(25)}$$
has yet to be investigated.

Having done the dimensional reduction and chosen the class of $A_{\z}$, the
way I propose looking at equations (1) is to regard (1a) and (1c) as
evolution equations for $A_z,A_{\z}$, and (1b) as a ``constraint'',
restricting some of the entries of $A_{\w}$; explicitly we have evolution
equations
$$ \eqalign{ \partial_wA_z  &= \partial_zA_w+[A_z,A_w] \cr
             \partial_wA_{\z}&=\partial_z A_{\w} - \partial_{\w}A_z
                            +[A_z,A_{\w}]+[A_{\z},A_w] \cr} \eqno{(26)}$$
where $\partial_{\w}A_{\z}=[A_{\z},A_{\w}]$ constrains certain entries
of $A_{\w}$ (we assume the class of $A_{\z}$ has been chosen so this has
a solution). Note these equations are invariant under the {\it full} group
of ($\z$-independent) gauge transformations. We obtain the restricted
notion of gauge invariance used in Drinfeld-Sokolov [15] and its
generalizations [26] when we partially fix the gauge by going to a gauge
where $A_{\z}$ is constant; there is still a residual gauge invariance
generated by the elements of the gauge algebra that commute with
the gauge-fixed $A_{\z}$. While this restricted notion of gauge invariance
allows us to understand some Miura maps, there are others that it cannot
explain. For example, NLS arises from (26) with
$$ \eqalign{
A_{\z}&=\pmatrix{\kappa&0\cr0&-\kappa\cr}~~~~~~~~~~~~~~~~~
A_{\w}=0\cr
A_z&=\pmatrix{0&\psi\cr\bar{\psi}&0\cr}  ~~~~~~~~~~~~~~~~~
A_w={\textstyle{1\over{2\kappa}}}
    \pmatrix{\partial_z^{-1}\partial_{\w}(\psi\bar{\psi})&
             \partial_{\w}\psi\cr
             \partial_{\w}\bar{\psi}&
             -\partial_z^{-1}\partial_{\w}(\psi\bar{\psi})\cr}
       } \eqno{(27)}$$
where $\kappa$ is a non-zero constant and  $\psi,\bar{\psi}$ are functions,
and the Heisenberg ferromagnet equation arises from (26) with
$$ \eqalign{
A_{\z}=\pmatrix{\lambda&\mu\cr\nu&-\lambda\cr}~~~~~~~~&~~~~~~~~~
A_{\w}=\pmatrix{\lambda F&\mu F+\mu_{\w}/2\lambda \cr
                \nu F-\nu_{\w}/2\lambda &-\lambda F\cr} \cr
A_z=&A_w=0 } \eqno{(28)}$$
where $\lambda,\mu,\nu$ are functions satisfying
$\lambda^2+\mu\nu=\kappa^2$ and $F$ is defined by $4\kappa^2F_z=
\mu(\lambda^{-1}\nu_{\w})_z -
\nu(\lambda^{-1}\mu_{\w})_z$. (In fact (27) and (28) give dimensional
deformations of the NLS and ferromagnet equations; the usual equations are
recovered by the further dimensional reduction $\partial_z=\partial_{\w}$.)
The gauge equivalence between NLS and the
Heisenberg ferromagnet, as given in [27], emerges from the full gauge freedom
of (26)\footnote*{For an analysis of many of the equations in the NLS gauge
equivalence class see [28]. There I defined the ``Ur-NLS'' equation
$$ \eqalign{S_w&={{2S_zT_{zz}}\over{T_z}}+3S_z^2-S_{zz}\cr
            T_w&=T_{zz}+2T_zS_z} $$
A solution of the $\kappa={1\over 2}$ ferromagnet equation (equation (52)) is
obtained from this by setting
$$ \eqalign{\lambda&={\textstyle{1\over 2}}+TS_z/T_z\cr
            \mu&=-T(1+TS_z/T_z)\cr
            \nu&=S_z/T_z } $$
In [29] Hirota showed the gauge equivalence of NLS and 	the ferromagnet
equation to a third equation (the equation in the abstract of [29]);
this is obtained from Ur-NLS, modulo some minor rescalings, via $\phi=-Te^S$,
$\bar{\phi}=S_ze^{-S}/T_z$.}.

There is a refinement of equations (26) we can make. We can solve the first
of equations (26) via equation (5a), and thus replace (26) with the system
$$ \eqalign{\partial_w g &= g A_w \cr
        \partial_wA_{\z}&=\partial_z A_{\w} - \partial_{\w}(g^{-1}\partial_z g)
            +[g^{-1}\partial_z g,A_{\w}]+[A_{\z},A_w] \cr} \eqno{(29)}$$
These equations have standard gauge invariance, i.e. invariance under
$A_{\w}\rightarrow\Lambda^{-1}A_{\w}\Lambda+\Lambda^{-1}\partial_{\w}\Lambda,~
A_{\z}\rightarrow\Lambda^{-1}A_{\z}\Lambda,~g\rightarrow g\Lambda$, but they
also have an {\it extra} invariance under $g\rightarrow tg$ where $t$ is a
element of the gauge
group\footnote{**}{Throughout this paper I have used the term ``gauge
group'' in the sense of physicists, as the structure group of the theory;
mathematicians use the term ``gauge group'' to refer to what I have called
the ``group of gauge transformations''. I point this out here so that it
should be clear that $t$ is a constant, unlike $\Lambda$ above, which is
dependent on the coordinates.}. Solutions of (29) related by
$g\rightarrow tg$ clearly give the same solution of (26). But
instead of passing directly from the variable $g$ to the variable
$A_z$ which is invariant under the whole group of $t$ transformations,
we could
pass first to some set of variables invariant under just a subgroup
of $t$ transformations; (29) will imply some evolution for these
intermediate variables. Using $t$ symmetry alone, from an integrable system
obtained from (29) we will obtain a whole series of integrable systems, by
``modding out'' by any subgroup of the gauge group. This explanation of
the relationship of the the various equations related to KdV, which all stem
{}from an equation called Ur-KdV with a $SL(2)$ invariance was first found
by Wilson [30]. We will derive Ur-KdV from (29) shortly; equations (29)
can also be used to derive the Ur-NLS equation of [28] (see the footnote
to the preceding paragraph).

While the general scheme outlined for reductions in the previous two
paragraphs seems to be correct, there is one subtlety; in explicitly
relating equations (26) or (29) to known integrable systems by writing
everything in components and simplifying, the need arises to fix certain
integration constants. It seems that the choice of integration
constants can be regarded as the fixing of certain gauge invariant quantity,
but I am not aware at the moment of a method of picking out the relevant
quantities\footnote{$\dagger$}{The problem here seems to arise because SDYM
typically contains many copies of any particular integrable system;
for example, the parameter $\kappa$ in (23), is a gauge invariant quantity,
and each value of $\kappa$ gives a reduction of SDYM to NLS.}.
To see that in spite of this the general scheme  we have explained is correct,
an example is now in order.

\vskip.2in

\noindent{\it 6. Example: KdV}
\smallskip
\noindent To illustrate the relation of equations (26) and (29), and the
issue of integration constants raised above,
let us look at (26) and (29) in the KdV case, with the partial gauge fixing
$$ A_{\z}=\pmatrix{0&0\cr1&0\cr} \eqno{(30)}$$
The second equations of both (26) and (29) become constraints. The most
general solution of the constraints for (26) yields
$$ A_z=\pmatrix{d&e\cr f&-d\cr} ~~~~~~~~
   A_w=\pmatrix{dj+{1\over 2}(f_{\w}-j_z)&ej-d_{\w}\cr c&
        -dj+{1\over 2}(j_z-f_{\w})\cr} ~~~~~~~~
   A_{\w}=\pmatrix{0&0\cr j&0\cr}  \eqno{(31)}$$
where $c,d,e,f,j$ are some functions of $z,w,\w$, with $e_{\w}=0$.
This last condition is where the integration constant becomes necessary;
we will take $e$ to be an arbitrary constant. A gauge invariant way of
saying this is that we will fix $Tr[A_{\z}(A_z-h^{-1}\partial_z h)]=
Tr[\partial_{\z}h h^{-1}\partial_{\z}N]$, but this is not very
illuminating. Residual gauge transformations act via
$$\eqalign{d&\rightarrow d+eu\cr
           f&\rightarrow f+u_z-2du-eu^2\cr
                          }\eqno{(32)}$$
where the function $u$ is the parameter of gauge transformations.
Fixing the gauge to $d=0$, the first equation of (26) now reduces to
the dimensional deformation of KdV mentioned in section 4:
$$ f_w= {\textstyle{1\over 2}}(-{\textstyle{1\over{2e}}}\partial_z^2
                               +f+\partial_zf\partial_z^{-1})f_{\w}
      \eqno{(33)}$$

Let us look at the corresponding reduction of (29). Writing
$$ g=\pmatrix{\alpha&\beta\cr \gamma&\delta\cr} \eqno{(34)}$$
where $\alpha\delta-\beta\gamma=1$, we have
$$ g^{-1}\partial_z g =
   \pmatrix{\delta\alpha_z-\beta\gamma_z &
            \delta\beta_z-\beta\delta_z  \cr
            \alpha\gamma_z-\gamma\alpha_z &
            \alpha\delta_z-\gamma\beta_z \cr} \eqno{(35)}$$
and it is no surprise that in the reduction we find we have to fix an
integration constant, which we do by setting $\beta\delta_z-\delta\beta_z=e$.
Residual gauge transformations act via
$$ \eqalign{\alpha&\rightarrow\alpha+u\beta\cr
            \gamma&\rightarrow\gamma+u\delta\cr}\eqno{(36)}$$
with $\beta,\delta$ invariant. We find that under gauge transformation
$\delta\alpha_z-\beta\gamma_z \rightarrow \delta\alpha_z-\beta\gamma_z + eu$
so the natural gauge choice is to set $\delta\alpha_z-\beta\gamma_z=0$.
Solving all these constraints on $g$ we eventually find
$$\eqalign{ g&=\pmatrix{\delta^{-1}+e^{-1}q\delta_z&
                        \delta q\cr e^{-1}\delta_z & \delta\cr} \cr
        g^{-1}\partial_z g&= \pmatrix{0&e\cr f&0\cr} \cr}  \eqno{(37)}$$
where $\delta=(eq_z^{-1})^{1\over 2}$ and $f=-{1\over{2e}}\{q;z\}$
($\{q;z\}$ denotes the Schwarzian derivative of $q$). With this choice of $g$
and
$$ A_w=\pmatrix{{1\over 4}\partial_{\w}f&
                {1\over 2}e\partial_z^{-1}\partial_{\w}f \cr
                {1\over 2}f\partial_z^{-1}\partial_{\w}f
                     -{1\over {4e}}\partial_z\partial_{\w}f&
                -{1\over 4}\partial_{\w}f&\cr}
   ~~~~~~~~
   A_{\w}=\pmatrix{0&0\cr {1\over 2}\partial_z^{-1}\partial_{\w}f &0\cr}
    \eqno{(38)}$$
we obtain from (29) the dimensional deformation of the Ur-KdV equation
$$ q_w=-{1\over{4e}}q_z\partial_z^{-1}\partial_{\w}\{q;z\}  \eqno{(39)}$$
It is easy to check directly that if $q$ solves (39)
$f\equiv-{1\over{2e}}\{q;z\}$ solves (33).

\vskip.2in

\noindent{\it 7. Some Formal Manipulations with Formulae of
     Drinfeld and Sokolov}
\smallskip
\noindent Having discussed issues of choice of gauge in reductions of
SDYM in section 5, we are almost ready to explain the origin of bihamiltonian
structure, but one more piece of groundwork is necessary. Even if it is known
how to obtain a certain integrable system from a larger one, and the larger
one has a known hamiltonian structure, if the hamiltonian structure
is presented as in section 4, via a hamiltonian operator ${\cal D}$,
or equivalently by
a set of Poisson brackets, it is an arduous procedure to reduce the
hamiltonian structure, via ``Dirac reduction''. As discussed in [30],
an alternative to writing a hamiltonian operator is to write a symplectic
form on the space of fields. If the hamiltonian operator is ${\cal D}$,
and the fields are arranged in a  column  vector ${\bf u}$, then the
associated form can be formally written
$\int dx \delta{\bf u}^T\wedge{\cal D}^{-1}\delta{\bf u}$.
If a form can be written then it is simple to perform a reduction by
pulling back the form to the reduced space of fields. Problems arise
though due to the need to invert the operator ${\cal D}$, which can not
always be done (there are systems for which the reverse is true, that the
operator ${\cal D}^{-1}$ is well-defined, and cannot be sensibly inverted).
In the context of bihamiltonian systems, as presented in section 4, there
are an infinite number of hamiltonian operators we can formally
construct, namely
${\cal R}^n{\cal D}_1=({\cal D}_2{\cal D}_1^{-1})^n{\cal D}_1$, where $n$
can be any integer and we might hope that certain ones of these are
``inverse local'', so the associated form can be written down.

As mentioned before, to obtain the $SL(N)$ Drinfeld-Sokolov [15] system
{}from SDYM we reduce by imposing $\partial_{\z}=0$ and
take $(A_{\z})_{ij}=\delta_{iN}\delta_{j1}$; we find from
the second equation of (26) that we need to choose integration constants,
which we can do by setting $([A_z]_+)_{ij}=\delta_{i+1,j}$, where $[M]_+$
denotes the strictly upper triangular part of the matrix $M$. The
Drinfeld-Sokolov systems describe evolution of such a matrix $A_z$, modulo
certain gauge transformations. Looking at the formulae for the
hamiltonian structures in section 3 of [15], we find that appropriate
formal expressions for the symplectic forms associated with the
first, second and third hamiltonian structures of the Drinfeld-Sokolov
systems are
$$\eqalign{
 \omega_1&=\int dz~ Tr (\delta A_z \wedge D_{\z}^{-1} \delta A_z) \cr
 \omega_2&=\int dz~ Tr (\delta A_z \wedge D_z^{-1} \delta A_z) \cr
 \omega_3&=\int dz~ Tr (\delta A_z \wedge D_z^{-1}D_{\z}D_z^{-1} \delta A_z) }
             \eqno{(40)}$$
Here $D_z=\partial_z+[A_z,~~],~D_{\z}=[A_{\z},~~]$; $D_{\z}^{-1}$
is at first glance much more problematic to define than $D_z^{-1}$, but in
the above formulae it fortunately only acts on lower triangular matrices.
Remarkably, there is a simple way to make two of the above forms well-defined;
if we write
$A_z=g^{-1}\partial_zg$ then $\delta A_z = D_z(g^{-1}\partial_z g)$,
so if we use the variable $g$ as fundamental we have
$$\eqalign{
 \omega_2&=-\int dz~ Tr (g^{-1}\delta g\wedge D_z(g^{-1}\delta g) ) \cr
 \omega_3&=\int dz~ Tr (g^{-1}\delta g\wedge D_{\z}(g^{-1}\delta g) ) \cr}
         \eqno{(41)}$$
While the manipulations in this section have been questionable, the result,
equation (41), is perfectly reasonable; it suggests that in terms of the
variables $g$ the Drinfeld-Sokolov systems have a well-defined
inverse-local bihamiltonian structure. The above manipulations for $\omega_2$
were discussed by Wilson [30].

\vskip.2in

\noindent{\it 8. The Bihamiltonian Structure Induced on Reductions of SDYM}
\smallskip
\noindent On the space of solutions of SDYM there are three natural
closed two forms we can write down
$$ \Omega^{i}=\int ~ Tr(\delta A\wedge\delta A) \wedge
                   \alpha^{i}~~~~~~~~~i=1,2,3  \eqno{(42)}$$
where $\alpha^i$, $i=1,2,3$ are three closed two-forms on ${\bf R}^4$
$$ \eqalign{\alpha^1&=dz\wedge dw  \cr
            \alpha^2&=d\z\wedge d\w \cr
            \alpha^3&=dz\wedge d\w - dw\wedge d\z \cr} \eqno{(43)}$$
These are natural, because assuming we are dealing with gauge fields that
behave well at infinity, the self-duality equations (1), which can be
written $F\wedge\alpha^i=0,~i=1,2,3$,  emerge as moment
maps for the gauge invariance of $\Omega^i$ [31]. We will consider these forms
under the circumstances that the fields are well-behaved in the $z,\z,\w$
directions, but possibly not in the $w$ direction (we could choose to
compactify in the $z,\z,\w$ directions). The criterion for gauge invariance
of $\Omega^i$ on a region $V$ of ${\bf R}^4$ under an infinitesimal gauge
transformation $A\rightarrow A+D\Phi$ is
$$ \int_{\partial V}  Tr(\Phi\delta A)\wedge\alpha^i =0 \eqno{(44)}$$
So for us, with the behavior of the fields as specified,
$\Omega^1$ is invariant, and the criteria for $\Omega^2$ and
$\Omega^3$ to be invariant are, respectively:
$$ \eqalign{
\int_{w=\infty} dzd\z d\w Tr(\Phi\delta A_z)-
\int_{w=-\infty} dzd\z d\w Tr(\Phi\delta A_z)
        &=0 \cr
\int_{w=\infty} dzd\z d\w Tr(\Phi\delta A_{\z})-
\int_{w=-\infty} dzd\z d\w Tr(\Phi\delta A_{\z})
       &=0 \cr}              \eqno{(45)}$$
So in particular $\Omega^2$ will be invariant if we restrict
to a set of solutions with fixed $A_z$, and $\Omega^3$ will be invariant if
we restrict to a set of solutions with fixed $A_{\z}$.

Using the representation of self-dual gauge fields given at the end of section
3 (equations (5a) and (13)), we can write
$$\eqalign{
\delta A_w &= D_w(g^{-1}\delta g)  \cr
\delta A_z &= D_z(g^{-1}\delta g)  \cr
\delta A_{\w} &= D_{\w}(g^{-1}\delta g)+ D_w(g^{-1}\delta Pg)  \cr
\delta A_{\z} &= D_{\z}(g^{-1}\delta g)+ D_z(g^{-1}\delta Pg)  \cr}
    \eqno{(46)}$$
Substituting these expressions into the formulae for $\Omega^2$ and $\Omega^3$,
it can be seen that the integrands are total derivatives,
and thus $\Omega^2$ and $\Omega^3$ naturally define two symplectic forms
on the space of $w$-independent functions
$$ \eqalign{
\tilde{\Omega}^2&=\int dzd\z d\w Tr(g^{-1}\delta g \wedge
    D_z(g^{-1}\delta g)) \cr
\tilde{\Omega}^3&=\int dzd\z d\w Tr(g^{-1}\delta g \wedge
   [D_{\z}(g^{-1}\delta g)+2D_z(g^{-1}\delta Pg)] \cr
   }\eqno{(47)} $$
If we restrict to a subspace of fields where $A_{\z}$ is constant, then
we can use the equation $\delta A_{\z}=0$ to eliminate the $\delta P$ term
in $\tilde{\Omega}^3$, and (up to overall normalizations) we obtain the
$\omega_2,\omega_3$ of section 7, equation (41), the Drinfeld-Sokolov
symplectic forms.

Note the following:

\noindent
\item{1.} The above derivation of the Drinfeld-Sokolov forms
{}from $\Omega^2,\Omega^3$ is free
of any ``formal'' manipulations.
\item{2.} I have taken care to
use the representation of the potentials in terms of $g$ and $P$, as opposed
to the other representations of section 3, because if we wish to consider
reductions setting $\partial_{\z}$ to zero, the representation in terms
of $g$ and $P$ is still good, while the others are not.
\item{3.} It is straightforward to check directly the gauge invariance
properties of $\tilde{\Omega}^2$,$\tilde{\Omega}^3$; we find their
invariance requires $\int dzd\z d\w Tr(\Phi \delta A_z)=0$ and
$\int dzd\z d\w Tr(\Phi \delta A_{\z})=0$ respectively, as we would
expect. Clearly the restricted gauge transformations of the Drinfeld-Sokolov
systems [15] satisfy these conditions.

We would also like to obtain hamiltonian structures for equations (26)
in an $A_z=0$ gauge. For this we use the standard $M$ formulation of
section 3, i.e. equations (5b) and (7). In light of note 2 above we should
be cautious. We have
$$\eqalign{
\delta A_{\w} &= D_{\w}(h^{-1}\delta h)  \cr
\delta A_{\z} &= D_{\z}(h^{-1}\delta h)  \cr
\delta A_w &= D_w(h^{-1}\delta h)+ D_{\w}(h^{-1}\delta Nh)  \cr
\delta A_z &= D_z(h^{-1}\delta h)+ D_{\z}(h^{-1}\delta Nh)  \cr
  }  \eqno{(48)}$$
These expressions allow us to write the integrands of $\Omega^1,\Omega^3$
as total derivatives, so we can write the associated forms
$\tilde{\Omega}^1$,$\tilde{\Omega}^3$ (the latter should presumably agree
with that above), and then restricting to the subspace $\delta A_z=0$
we get
$$ \eqalign{ \tilde{\Omega}^1&=\int dzd\z d\w Tr (h^{-1}\delta h \wedge
                  D_{\z}(h^{-1}\delta h)) \cr
\tilde{\Omega}^3&=\int dzd\z d\w Tr (h^{-1}\delta h \wedge
                  D_z(h^{-1}\delta h)) \cr }\eqno{(49)}$$
This is as far as we can go without formal manipulations; but by analog
with the successful manipulations of the previous section it is natural
to guess that for reductions of SDYM in $A_z=0$ gauge, hamiltonian structures
arise from the following symplectic forms on the space of potentials $A_{\z}$:
$$\eqalign{
 \tilde{\omega}_1&=\int dz~
          Tr (\delta A_{\z} \wedge D_z^{-1} \delta A_{\z}) \cr
 \tilde{\omega}_2&=\int dz~
          Tr (\delta A_{\z} \wedge D_{\z}^{-1} \delta A_{\z}) \cr
 \tilde{\omega}_3&=\int dz~
          Tr (\delta A_{\z} \wedge D_{\z}^{-1}D_zD_{\z}^{-1} \delta A_{\z}) }
             \eqno{(50)}$$
$\tilde{\omega}_2$ and $\tilde{\omega}_3$ give $\tilde{\Omega}_1$ and
$\tilde{\Omega}_3$ above, and $\tilde{\omega}_1$ is motivated by the usual
recursion formalism. In equation (50) $D_z=\partial_z$ and
$D_{\z}=[A_{\z},~~]$. In at least the simplest example, the Heisenberg
ferromagnet, it seems these formulae have some meaning.

\vskip.2in

\noindent{\it 9. Example: The Heisenberg Ferromagnet}
\smallskip
\noindent In equation (28) the method of reduction of SDYM to the
Heisenberg ferromagnet  was given. It is straightforward to see
that $\tilde{\omega}_2$ corresponds to the standard Poisson
brackets of [27]. Here I wish to look at $\tilde{\omega}_1$. Since
$\lambda^2+\mu\nu=\kappa^2$, we can eliminate $\delta\lambda$ from
$\tilde{\omega}_1$ and write it (up to an overall normalization)
$$ \int dz~\pmatrix{\delta\mu & \delta\nu\cr} \wedge
   \pmatrix{{{\nu}\over{\lambda}}\partial_z^{-1}{{\nu}\over{\lambda}}&
            2\partial_z^{-1}+
            {{\nu}\over{\lambda}}\partial_z^{-1}{{\mu}\over{\lambda}}\cr
            2\partial_z^{-1}+
            {{\mu}\over{\lambda}}\partial_z^{-1}{{\nu}\over{\lambda}}&
            {{\mu}\over{\lambda}}\partial_z^{-1}{{\mu}\over{\lambda}}\cr}
    \pmatrix{\delta\mu \cr \delta\nu\cr} \eqno{(51)}$$
The ferromagnet equations are
$$\eqalign{
2\kappa^2\mu_w &=\partial_z(\lambda\mu_z-\mu\lambda_z)\cr
-2\kappa^2\nu_w&=\partial_z(\lambda\nu_z-\nu\lambda_z)\cr}
            \eqno{(52)}$$
The above symplectic form  gives a hamiltonian structure since we can write:
$$ \eqalign{
  \pmatrix{{{\nu}\over{\lambda}}\partial_z^{-1}{{\nu}\over{\lambda}}&
            2\partial_z^{-1}+
            {{\nu}\over{\lambda}}\partial_z^{-1}{{\mu}\over{\lambda}}\cr
            2\partial_z^{-1}+
            {{\mu}\over{\lambda}}\partial_z^{-1}{{\nu}\over{\lambda}}&
            {{\mu}\over{\lambda}}\partial_z^{-1}{{\mu}\over{\lambda}}\cr}
   \pmatrix{\partial_z(\lambda\mu_z-\mu\lambda_z)\cr
            -\partial_z(\lambda\nu_z+\nu\lambda_z)\cr}
   &={{2\kappa^2}\over{\lambda}}\pmatrix{-\nu_z\cr\mu_z\cr} \cr
   &=2\kappa^2\pmatrix{\delta/\delta\mu \cr
             \delta/\delta\nu \cr}
    \int dz~{{\nu\mu_z-\mu\nu_z}\over{\lambda+\kappa}}  }\eqno{(53)}$$
The hamiltonian in the above equation was given in [27]; I am uncertain
whether the ``group theoretic origin'' of this hamiltonian structure
for the ferromagnet equation has been realised before.

\vskip.2in

\noindent{\it 10. The KP Hierarchy from the SDYM Hierarchy}\footnote{
$\dagger\dagger$}{The results in this section were obtained in collaboration
with Didier Depireux. Simliar results have been obtained by I.A.B.Strachan.}
\smallskip
\noindent Let us choose a gauge algebra ${\cal G}$ such that a) ${\cal G}$
can be identified with its universal enveloping algebra, i.e. we can
``multiply'' two elements of ${\cal G}$ to get another element of ${\cal G}$,
and b) ${\cal G}$ ``splits'', i.e. we can write ${\cal G}={\cal G}_+\oplus
{\cal G}_-$, where the commutator of two elements of ${\cal G}_+$ is in
${\cal G}_+$, and the commutator of two elements of ${\cal G}_-$ is in
${\cal G}_-$. For $M\in{\cal G}$ define $M_+,M_-$ as the projections
of $M$ onto ${\cal G}_+$ and ${\cal G}_-$ respectively.

Suppose $L\in{\cal G}$ and consider the following evolution equations for $L$:
$$ {{\partial L}\over{\partial z_i}}=[(L^i)_+,L] ~~~~~~~~~i=1,2,....
             \eqno{(54)}$$
These imply (see section 4 of [32])
$$\eqalign{
{{\partial L^j}\over{\partial z_i}}&=[(L^i)_+,L^j] \cr
{{\partial (L^j)_+}\over{\partial z_i}} -
     {{\partial (L^i)_+}\over{\partial z_j}} &= [(L^i)_+,(L^j)_+] \cr
{{\partial (L^j)_-}\over{\partial z_i}} -
     {{\partial (L^i)_-}\over{\partial z_j}} &= -[(L^i)_-,(L^j)_-] \cr}
      \eqno{(55)}$$
It is straightforward to check, using these results, that equations (54)
for $L$ are equivalent to the equations of the SDYM hierarchy (4) with a
dimensional reduction $\partial_{\z_i}=\partial_{z_{i-1}}$, $i=1,2,...$
(so $\partial_{\z_1}=0$) and an ansatz
$$ \eqalign{
A_{\z_1}&=L \cr
A_{\z_i}&=(L^{i-1})_-,~~~~~~~~~i>1 \cr
A_{z_i} &=-(L^i)_+,   ~~~~~~~~~~~i\ge 1\cr} \eqno{(56)}$$

The KP hierarchy is just this with ${\cal G}_+$ the algebra of finite order
differential operators in one variable $x$, ${\cal G}_-$ the algebra of
psuedodifferential operators in $x$ (with no ``constant term''), and
$$ L=\partial_x+u_2(x)\partial_x^{-1}+u_3(x)\partial_x^{-1}+....
         \eqno{(57)}$$
The gauge choice here is one in which $A_{z_1}=\partial_x$;
in the Sato theory (see for example [32]) an object $W$ is defined
such that $L=W\partial_xW^{-1}$, and this defines a gauge transformation
to a gauge where $A_{\z_1}=\partial_x$.

\vskip.2in

\noindent{\it 11. Conclusions and Further Directions}
\smallskip
\noindent That the bihamiltonian structures of integrable systems arise
naturally from structures on the space of solutions of SDYM is a most
pleasing result, and lends much support to the idea that the interesting
properties of integrable systems all stem from SDYM. But there is a long
way to go before the theory of integrable systems can be rewritten from
this viewpoint. Many interesting questions remain, of which I will just pose
three here:

\noindent
\item{1)} It is interesting that we recover most standard integrable
equations and their hierarchies by a reduction of SDYM and its hierarchy
by essentially half the number of coordinates. One the other hand, there
are interesting approaches in the theory of integrable systems which
involve adding extra coordinates; examples are the mysterious Hirota bilinear
operators [29], and the unified approach to integrable systems of Fokas
and Santini [33]. It would be very interesting if a link could be made
between the auxiliary coordinates in these methods, and the coordinates
we have reduced by from SDYM.
\item{2)} The fact that we now have a simple reduction of SDYM to KP
opens a lot of possibilities. First, using the methods of this paper,
we should be able to extract the bihamiltonian structure of KP [34], and
understand hierarchies related to KP via gauge transformations, such as
the modified KP. These issues are currently under investigation
in collaboration with D.Depireux. More importantly, we need to
understand how the twistor formalism gives inverse scattering formalism
(and hopefully the Hirota method of direct solution) for KP and its
relatives (Davey-Stewartson, multi-component KPs etc.).
\item{3)} Quantized integrable systems have been extensively studied,
and arise as deformations of conformal field theories. It would be
interesting too see if there exists a quantization of SDYM that unifies
quantized integrable systems in some sense. Quantization of SDYM using
the lagrangians of [19] and [20] is problematic, as there seem to be
renormalization problems, but it seems the $N=2$ string can be regarded in
some sense as a quantization of SDYM [35], and this might give positive
results.

\vskip.2in

\noindent{\it 12. Acknowledgements}
\smallskip
\noindent
I thank D.Depireux, O.Lechtenfeld and V.P.Nair for discussions.
This work was supported by a grant in aid from the U.S.Department of
Energy, \#DE-FG02-90ER40542.

\vskip.2in

\noindent{\it 13. References}
\smallskip
\noindent
\item{[1]} E.Witten, {\it Phys.Rev.Lett.} {\bf 38} (1977) 121;
 A.N.Leznov and M.V.Savelev, {\it Comm.Math.Phys.} {\bf 74} (1980) 111.
\item{[2]} R.S.Ward, {\it Phil.Trans.Roy.Soc.Lond.A} {\bf 315} (1985) 451,
 and in {\it Field Theory, Quantum Gravity and Strings}, ed. H.J.de Vega
 and N.Sanchez, Springer Lecture Notes in Physics Vol. 246 (1986).
\item{[3]} L.J.Mason and G.A.J.Sparling {\it Phys.Lett.A} {\bf 137} (1989) 29,
  {\it J.Geom.Phys.} {\bf 8} (1992) 243.
\item{[4]}  I.Bakas and D.A.Depireux, {\it Mod.Phys.Lett.} {\bf A6} (1991)
    399, {\it Int.J.Mod.Phys.} {\bf A7} (1992) 1767, {\it Mod.Phys.Lett.}
   {\bf A6} (1991) 1561 (erratum {\it ibid.} {\bf A6} (1991) 2351).
\item{[5]} J.Schiff, in {\it Painlev\'e Transcendents: Their Asymptotics
  and Applications}, ed. D.Levi and P.Winternitz, Plenum (1992).
\item{[6]} I.A.B.Strachan, ``Some Integrable Hierarchies in (2+1) Dimensions
 and their Twistor Description'', Oxford preprint to appear in
 {\it J.Math.Phys.};
 ``Null Reductions of the Yang-Mills self-duality Equations and Integrable
  Models in (2+1) Dimensions'', talk given at the NATO workshop, Exeter,
  England (July 1992).
\item{[7]} M.J.Ablowitz, S.Chakravarty and L.A.Takhtajan, ``A Self-Dual
   Yang-Mills Hierarchy and Its Reductions to Integrable Systems in 1+1
   and 2+1 Dimensions'', Colorado preprint, August 1992.
\item{[8]} A.A.Belavin and V.E.Zakharov, {\it Phys.Lett.B} {\bf 73} (1978) 53.
\item{[9]} J.Weiss, M.Tabor and G.Carnevale, {\it J.Math.Phys.} {\bf 24} (1983)
   522.
\item{[10]} M.Jimbo, M.D.Kruskal and T.Miwa, {\it Phys.Lett.A} {\bf 92}
   (1982) 59; R.S.Ward {\it Phys.Lett.A} {\bf 102} (1984) 279;
    {\it Nonlinearity} {\bf 1} (1988) 671.
\item{[11]} E.Corrigan, D.B.Fairlie, R.G.Yates and P.Goddard,
   {\it Phys.Lett.B} {\bf 72} (1978) 354; {\it Comm. Math.Phys.} {\bf 58}
   (1978) 223.
\item{[12]} L.Crane, {\it Comm.Math.Phys.} {\bf 110} (1987) 391.
\item{[13]} B.Grammaticos, A.Ramani and J.Hietarinta, {\it J.Math.Phys.}
   {\bf 31} (1990) 2572.
\item{[14]} T.J.Hollowood and J.L.Miramontes, ``Tau-functions and Generalized
   Integrable Hierarchies'', Oxford/CERN preprint OUTP-92-15P,
   CERN-TH-6594/92, hep-th/9208058.
\item{[15]} V.G.Drinfeld and V.V.Sokolov, {\it Jour.Sov.Math.} {\bf 30}
    (1985) 1975.
\item{[16]} R.S.Ward, {\it Nucl.Phys.B} {\bf 236} (1984) 381.
\item{[17]} C.N.Yang, {\it Phys.Rev.Lett.} {\bf 38} (1977) 1377.
\item{[18]} M.Bruschi, D.Levi and O.Ragnisco, {\it Lett.Nuov.Cim} {\bf 33}
   (1982) 263.
\item{[19]} A.N.Leznov and M.A.Mukhtarov, {\it J.Math.Phys.} {\bf 28} (1987)
  2574; A.Parkes, {\it Phys.Lett.B} {\bf 286} (1992) 265.
\item{[20]} V.P.Nair and J.Schiff, {\it Nucl.Phys.} {\bf B371} (1992) 329.
\item{[21]} K.Takasaki, {\it Comm.Math.Phys.}{\bf 127} (1990) 225.
\item{[22]} P.J.Olver, {\it Applications of Lie Groups to Differential
   Equations}, Springer-Verlag (1986).
\item{[23]} F.Calogero, {\it Lett.Nuov.Cim.} {\bf 14} (1975) 443;
     P.Santini, ``Algebraic Properties and Symmetries of Integrable
    Evolution Equations'', La Sapienza preprint (1988).
\item{[24]} E.Witten, {\it Comm.Math.Phys.} {\bf 92} (1984) 455.
\item{[25]} A.P.Fordy and P.P.Kulish {\it Comm.Math.Phys.} {\bf 89} (1983)
 427; see also I.A.B.Strachan, {\it J.Math.Phys.} {\bf 33} (1992) 2477.
\item{[26]} M.F.de Groot, T.J.Hollowood and J.L.Miramontes,
   {\it Comm.Math.Phys.} {\bf 145} (1992) 57; N.J.Burroughs,
   M.F.de Groot, T.J.Hollowood and J.L.Miramontes, {\it Phys.Lett.B}
  {\bf 277} (1992) 89, and  ``Generalized Drinfeld-Sokolov Hierarchies 2:
   The Hamiltonian Structures'', Institute for
    Advanced Study/ Princeton University preprint IASSNS-HEP-91/19,
    PUPT-1251, hep-th/9109014.
\item{[27]} V.E.Zakharov and L.A.Takhtadzhyan, {\it Theor.Math.Phys.}
   {\bf 38} (1979) 17.
\item{[28]} J.Schiff, ``The Nonlinear Schr\"odinger Equation and Conserved
   Quantities in the Deformed Parafermion and SL(2)/U(1) Coset Models'',
   Institute for Advanced Study preprint IASSNS-HEP-92-57, hep-th/9210029.
\item{[29]} R.Hirota, in {\it Non-linear Integrable Systems - Classical
   Theory and Quantum Theory}, ed. M.Jimbo and T.Miwa, World Scientific
   (1981).
\item{[30]} G.Wilson, {\it Phys.Lett.A} {\bf 132} (1988) 45,
       {\it Quart.J.Math.Oxford} {\bf 42} (1991) 227,
      {\it Nonlinearity} {\bf 5} (1992) 109, and in {\it Hamiltonian
     Systems, Transformation Groups and Spectral Transform Methods},
     ed. J.Harnad and J.E.Marsden, CRM (1990).
\item{[31]} N.Hitchin, {\it Monopoles, Minimal Surfaces and Algebraic Curves},
   S\'eminaire de Math\'ematiques Sup\'erieures Volume 105, Les Presses
   de l'Universit\'e de Montr\'eal (1987).
\item{[32]} Y.Ohta, J.Satsuma, D.Takahashi and T.Tokihiro,
  {\it Prog.Th.Phys.Suppl.} {\bf 94} (1988) 210.
\item{[33]} A.S.Fokas and P.Santini, in {\it Solitons in Physics, Mathematics
  and Nonlinear Optics}, ed. P.J.Olver and D.H.Sattinger, Springer-Verlag
  (1990).
\item{[34]} L.A. Dickey, {\it Soliton equations and Hamiltonian systems},
   World Scientific, 1991.
\item{[35]} C.Vafa and H.Ooguri,
   {\it Mod.Phys.Lett.} {\bf A5} (1990) 1389,
   {\it Nucl.Phys.B} {\bf 367} (1991) 83,
   {\it Nucl.Phys.B} {\bf 361} (1991) 469.

\bye